\documentclass[preprint,12pt]{elsarticle}

\usepackage{amssymb}
\journal{International Journal Numerical methods for Heat \& Fluid flow}

\begin{document}

\begin{frontmatter}



\title{Dependence of the separative power of an optimised Iguassu gas centrifuge on the velocity of rotor}


\author{S.V. Bogovalov, V.D. Borman, V.D.Borisevich, I.V.Tronin, V.N.Tronin }
\ead{svbogovalov@mephi.ru}
\address{National research nuclear university (MEPHI), Kashirskoje shosse, 31, 115409, Moscow, Russia}

\begin{abstract}
{\bf Purpose.} The objective of the work is to determine dependence of the separative power of the optimised Iguassu gas centrifuge on the velocity of the rotor. \\
{\bf Methodolgy.} The dependence is determined by means of computer simulation of the gas flow in the gas centrifuge and numerical solution
of the diffusion equation for the light component of the binary mixture of uranium isotopes. 2-D axisymmetric model with the sources/sinks of the mass, angular momentum and 
energy reproducing the affect of the scoops was explored for the computer simulation.  Parameters of the model correspond to the parameters of the so called Iguassu centrifuge. The separative power has been
optimised in relation to the pressure of the gas, temperature of the gas, the temperature drop along the rotor,  power of the source of angular momentum and energy, feed flow and geometry of the lower baffle. In the result the optimised separative power depends only on the velocity, length and diameter of the rotor. \\
{\bf Findings.} The dependence on the velocity is described by the power law function with the power law index $2.6$ which demonstrate stronger dependence on the velocity than it follows from experimental data. However, the separative power obtained with limitation on the pressure  depends on the velocity on the power $\approx 2$ which well agree with the experiments. \\
{\bf Originality. } For the first time the optimised separative power of the gas centrifuges have been calculated via numerical simulation of the gas flow and diffusion of the binary mixture of the isotopes.

\end{abstract}


\end{frontmatter}


\section{Introduction}
\label{s1}

Knowledge of dependence of the separative power on the parameters of the gas centrifuge  (hereafter GC) is necessary for design of efficient GC for industrial  production of enriched uranium. Up to now this  knowledge  has basically an empirical character. In spite of 60-th year history of exploration of the gas centrifuges for enriched uranium production, there is no fundamental understanding of the dependence of the separative power of the centrifuges on the parameters. 

The upper limit on the separative power of GC has been estimated firstly by Dirac  \cite{cohen}. The separative power $\delta U_{max}$ of any GC is limited by the value
\begin{equation}
 \delta U_{max}={\pi \rho DL\over 2} \left( {\Delta M V^2\over 2RT} \right)^2,   
\label{eq1}
\end{equation}
where $\rho D$ is the density of uranium hexafluoride ($UF_6$) times the coefficient of self-diffusion of uranium isotopes $^{238}U$ and $^{235}U$. $\Delta M$ is the mass difference of molecules with  different uranium isotopes , $R$ is the  gas-law constant, $T$ is gas temperature, $L$ is the length of the rotor of GC, $V$ is the linear velocity of the rotor rotation. 
 
We are interested here in the separative power of GC optimised on all parameters which can be controlled by a designer. 
The separative power can be presented as a function of a lot of parameters $\delta U(V,L,T,a, \alpha_1, \alpha_2,...)$, where $a$ is the rotor diameter and the series of parameters $\alpha_i$ includes, for example, pressure at the wall of the rotor, feed flow $F$, feed cut $\theta$, variation of temperature along the rotor  $\delta T$ and many others.  Optimization of the gas centrifuge means a search for the maximum of this function at the variation of all the parameters $\alpha_i$.  Such a search is performed for every series of $V,~L,~T,~a$. Therefore, the separative power of the optimised GC depends only on the  limited set of the parameters $V,~L,~T,~a$.  Such a formulation of the problem carries additional difficulties in the solution of the problem because it is necessary not only to calculate the separative power of the GC, but additionally to optimise (to find maximal value) in relation to all possible parameters at fixed $V$, $L$, $a$ and $T$.

At the beginning of 1960's an Onsager group from US developed a theory called  ‘‘the pancake approximation’’  \cite{onsager,du}.  This approach gave the following equation for the optimised separative power
\begin{equation}
 \delta U= (0.038 V-11.5)L, ~kg\cdot SWU/yr.
\label{eq2}
\end{equation}
In contrast to eq. (\ref{eq1}) where the separative power increases as  $V^4$, in eq.( \ref{eq2}) the separative power grows linearly with $V$. This difference is crucially important for the gas centrifuge designers because equations (\ref{eq1}) and (\ref{eq2}) give essentially different predictions for the separative power at the growth of the rotor velocity. 

Experimental data, collected in Russia \cite{senchenko} gave   the following empirical equation
\begin{equation}
 \delta U=12L \left({V\over 700 ~{\rm m/s}}\right)^2\left({2a\over 12~{\rm cm}}\right)^{0.4}, ~kg \cdot SWU/yr,
\label{eq3}
\end{equation}
where $L$ is measured in meters. Recently this result has been well confirmed  by more extended experimental data \cite{yatsenko}.
%$\delta U \sim V^2$ in the empirical eq. (\ref{eq3}). 

After that, a new equation defining the separative power of GC has been  proposed in \cite{kemp} 
\begin{equation}
 \delta U= \left({V^2 L\over 33000}\right) e_E,~ kg\cdot SWU/yr,
\label{eq6}
\end{equation}
where $V$ is measured in $\rm m/s$, $L$ is the rotor length in meters, $e_E$ is a dimensionless experimental efficiency  which take into account specific features of every centrifuge and varies in the limit 0.4 -1.2. This equation  correctly reproduces the empirical law (\ref{eq3}). However, since this equation has been obtained in the pancake model which does not take into account important features of the real flow in the GC the question about dependence of the optimised separative power of the GC on the parameters remains an open problem up to now. 

Solution of this problem is important from practical point of view indeed. Simple estimates show that the maximal  separative power defined by equation (\ref{eq1}) is 4-5 times higher than the optimal separative power (\ref{eq3}) determined experimentally at $V=700~ \rm m/s$ and $2a=12~ \rm cm$. This remarkable difference is due to the different dependence of the separative power on $V$. Our final objective is to answer on  a few fundamental questions. What are the physical reasons for $V^2$ dependence in (\ref{eq3})? What factors limit the growth of $\delta U$ with $V$? Is it possible to dispose these factors and to increase the separative power of the gas centrifuges a few times at the same velocity and length of the rotor? In other words, is it possible to design a gas centrifuge a few times more efficient compared to the existing ones?  

Recently, one more step in understanding of the dependence of the separative power on the parameters has been done in \cite{bog2016}. A simplified concurrent GC has been considered in this work. The optimised separative power of such a GC is defined by the equation
\begin{equation}
 \delta U = 12.7 \left({V\over 700~ \rm m/s}\right)^2 \left({300~{\rm K}\over T}\right) L , ~\rm kg\cdot  SWU/year
\label{eq59}
\end{equation}
which well agrees with the empirical equation including the numerical coefficient. At first glance this result solves the problem. However, eq. (\ref{eq59}) has been obtained for the simplified concurrent centrifuge while the  centrifuges  used in industry are countercurrent. An axial countercurrent circulation of gas is excited in these centrifuges specially to increase their efficiency. Therefore, it is not evident that the separative power of the countercurrent centrifuges can be described by the same equation as the concurrent centrifuges.

The problem of the separative power of the concurrent centrifuges has been solved analytically. In the case of countercurrent centrifuges it is necessary to perform huge amount of computational work on numerical simulation and optimization of the GC. 
In this work for the first time we present results of calculations of the optimised  separative power of the countercurrent GC as a function of velocity. 

The paper is organised as follows. In the second section, we present the scheme of the countercurrent centrifuge, basic equations and assumptions.  In sec. 3 the solution is described in details. In sec. 4 the optimised separative power is calculated and finally, we discuss the solution in  last sec.5.

\section{2-D model of the Iguassu gas centrifuge}

\begin{figure}[h!]
 \centering
  \includegraphics[width=0.4\textwidth]{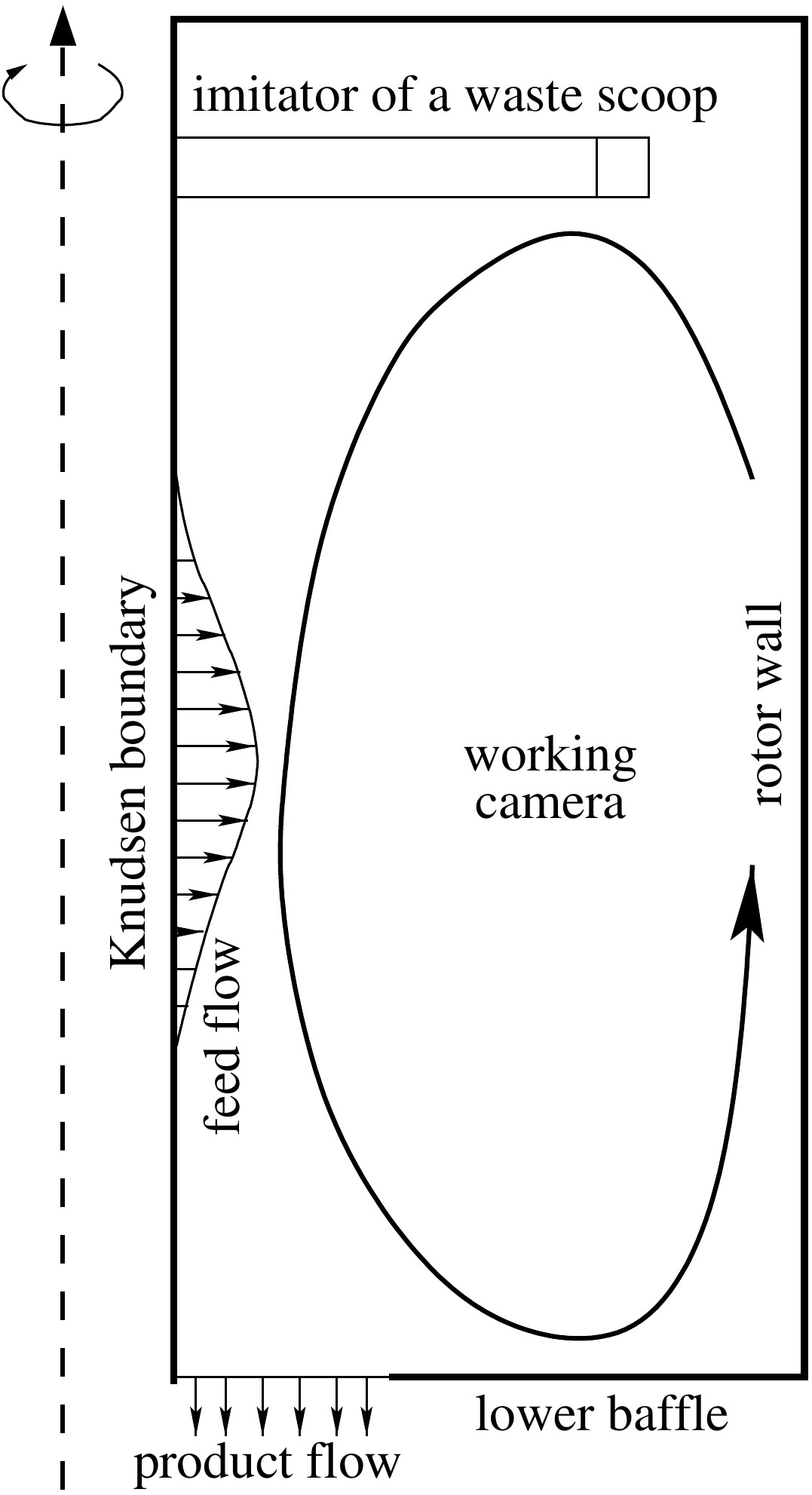}
  \caption{Scheme of the computational domain}
\label{fig1}
\end{figure}

The model Iguassu centrifuge \cite{iguasu} is widely explored for the numerical simulations. 
The modelling of the flow and separation is performed in the rotating frame system. The diameter of the rotor of the GC is $d = 0.12~ \rm m$. The length of the rotor  $L = 1 ~ \rm m$. The modelling is performed in axisymmetric approximation. The affect of the waste scoop has been modelled as a source of mass, momentum and energy distributed over a local toroidal region.  The computational domain consists of the working chamber and waste chamber located between two coaxial cylinders as it is shown in fig. \ref{fig1}. The outer cylinder corresponds to the rotor wall. The inner cylinder is the artificial wall introduced into the model to avoid the simulation in the rarefied gas where the hydrodynamical approximation is not valid \cite{knudsen}.
The product chamber located below the working chamber was not included into the computational domain. Its influence was modelled by the pressure imposed at the rotor wall near the lower baffle of the working chamber. The product flow  has been specified as a fraction $\theta$ of the feed flow. 

A linear temperature profile has been specified at the wall of the rotor. The minimal temperature $T_0$ has been specified above the temperature of sublimation of the working gas. At every step of the optimization procedure of the GC the temperature of  the gas has been calculated at the pressure exceeding on $15\%$ the current pressure at the wall. This temperature has been taken as $T_0$. The sublimation of the gas is avoided due to this procedure.

Internal boundary of the computational domain corresponds to the Knudsen zone were the gas pressure is of the order of 1 Pa. The feed flow has been specified at the surface of this boundary. 

\section{Numerical solution}
\subsection{Basic equations and numerical methods}

The solution of the problem of the optimised separation power on the parameters consists on two parts. First we have to solve full system of hydrodynamic equations defining the flow of the working gas with  molar mass
 $M$ and rotating with the angular velocity  $\omega$. The system of equations defining the dynamics of the gas in the rotating frame system is as follows  \citep{landau}:
\begin{equation}
 \frac{\partial \rho}{\partial t}+{\partial \rho { v_k}\over\partial { x_k}}=0,
\label{eq26}
\end{equation}
\begin{equation}
 \rho \frac{\partial { v}_i}{\partial t}+\rho \left( \frac{v_k\partial v_i }{\partial { x}_k}-\omega^2  r_i -2 \varepsilon_{iml}\omega_m v_l\right)=-\frac{\partial P}{\partial { x_i}}+\frac{\partial }{\partial x_k} \mu \left( \frac{\partial v_i}{\partial x_k}+ \frac{1}{3}\frac{\partial v_k}{\partial x_i} \right).
\end{equation}

\begin{eqnarray}
\frac{\partial \rho \left( c_v T+v^2/2-\omega^2r^2/2 \right)}{\partial t}=
\frac{\partial }{\partial x_k} ( \rho v_k \left( c_p T+v^2/2 -\omega^2r^2/2 \right) + &\nonumber\\
+\chi \frac{\partial T}{\partial x_k}+v_i(\frac{\partial v_i}{ \partial x_k}+\frac{\partial v_k}{\partial x_i}-\frac{1}{3}\delta_{ik}\frac{\partial v_l}{\partial x_l})) 
\label{eq5}
\end{eqnarray}
%\end{equation}
{ where $r_i$ are the components of the cylindrical radius, $c_p$ and $c_v$  are  the specific heat capacity for constant pressure and volume, $P$- pressure, $\rho$ - density, $T$ - temperature  and $v_i$  are the components of the velocity,  $\mu$  dynamic viscosity and $\chi$ is the thermal conductivity. }

These equations are supplemented by the equation for concentration $C$ of the $UF_6$ with light isotope of uranium $^{235}U$.   This equation has a form
\begin{equation}
 {\partial \rho C\over \partial t}
 -{\partial \over \partial x_i}\left(\rho v_i C-\rho D\left({\partial C\over \partial x_i}+{\Delta M\over M}C(1-C){\partial \ln{p}\over \partial x_i}\right)\right)=0.
\end{equation}

A specialised numerical codes has been developed in National research nuclear university (MEPHI) for solution of the full system of equations. The code is based on Godunov numerical scheme of the second order \cite{godunov}. The numerical solution has been obtained on the computer cluster of the university. 

\subsection{Verification  of the numerical code}

The codes for numerical simulation of the hydrodynamical flows and diffusion of the light isotope in the gas centrifuge need special procedures of verification. The verification of our numerical code have been performed using the methodology developed in \cite{bog13,kislov13}.
 
\section{Optimization of the separative power of the gas centrifuge.}
The second step to be performed consists in optimizing the GC over the whole range of parameters. The separative power $\delta U$ is calculated according to the well known equation proposed by Peierls \cite{cohen}
\begin{equation}
 \delta U= F_pV(C_p)+F_wV(C_w)-F V(C_f),
\end{equation}
where $F_p$ is the product flow of the gas enriched by uranium $^{235}U$, $C_p$ is the concentration of the light component in the product flow, $F_w$ is the waste flow,
$C_w$ is the concentration of the light component in the waste flow and $F$, $C_f$ are the feed flow and concentrations in the feed flow.
The target of the optimization procedure was to reach maximum of $\delta U$ at variation of the parameters of the gas centrifuge  with some limitations on these parameters.

The variable parameters were the pressure of the gas at the wall of the rotor $p_w$, temperature of the lower cap end of the rotor $T_0$, temperature variation between the upper  and lower cap ends of the rotor $\Delta T$, feed flow $F$, the breaking capacity of the waste scoop $W$ and the radius $r_b$ of the product baffle. In addition we varied the radius $R_{low}$ of the hole of the lower baffle for a total 7 parameters. Finally, the optimised separative power depends only on the length  $L$, velocity  $V$, radius  of the rotor $a$  and the feed cut $\theta$. 

$UF_6$ gas physical limitations must be considered when optimizing the different parameters. According to the equation of state, the gas goes through a sublimation point and forms a solid phase at the specified pressure and temperature. The sublimation is defined by the pressure and temperature. At a specified pressure there is a minimal temperature $T_{sub}(p)$. Above this temperature $UF_6$ exists in the gas phase. Thus we imposed the limitation that $p<0.85 \cdot P_{sub}(T)$, where $P_{sub}(T)$ is the pressure of sublimation at temperature $T$.

The BOBYQA~\cite{BOBYQA} direct search method within the NLopt package~\cite{nlopt} has been used for the optimization of the centrifuge. Optimization provides us the values of pressure at the wall of the rotor $p$, feed flow $F$, temperature drop along the wall of the rotor $\Delta T$  and power of the braking of the gas by the scoop $W$ at which the maximal value of the separative power of the GC is achieved.

All the calculations were performed in parallel using computer cluster. Single evaluation of the GC separative power uses 8 processor cores and takes from 10 to 50 minutes depending on the parameters of the GC. It is nessesary to perform from 10 to 100 evaluations in order to the BOBYQA optimization method converges. Therefore, one optimizations takes from several hours to several days to converge. Optimization calculations with different rotor velocities were done in parallel.

\section{Results}

\subsection{Dependence of the optimised separative power on the rotor velocity}

\begin{figure}[h!]
 \centering
  \includegraphics[width=0.8\textwidth]{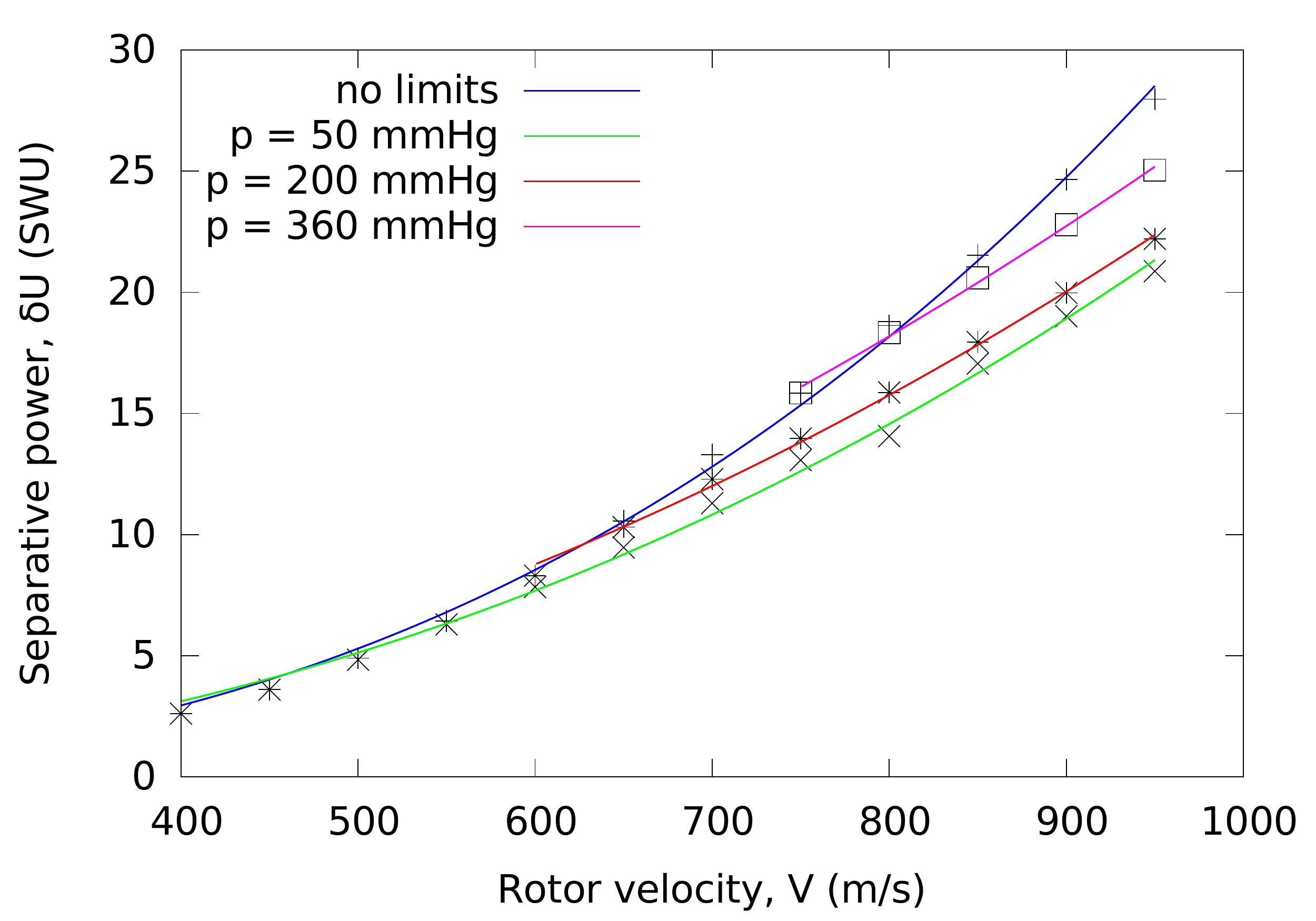}
  \caption{Dependence of the optimised separative power on the velocity of rotor at the conventional procedure of optimization (blue line). Dependence of the optimised separative power at the fixed pressure are shown in green ($p=50 ~\rm mm Hg$), red ($p=200 ~\rm mm Hg$) and magenta ($p=360~\rm  mm Hg$) lines.  }
\label{fig2}
\end{figure}

Dependence of the optimised separative power on the velocity of the Iguassu centrifuge is shown in fig. \ref{fig2}. The calculated dependence (crosses) is well described by the power law function with the index $\alpha = 2.62\pm 0.04$.
This dependence does not agree with the experimental measurements obtained for Russian centrifuges which gives the power law index $\alpha =2$. Analysis shows that the difference between our calculations and the experiment arises because of limitations on pressure. We did not impose any limitations on pressure at the walls of the rotor. In real experiments the pressure is always limited from above. Therefore, we have performed calculations of the dependence of the optimised separative power on velocity for  3 different fixed pressures in the rotor. These dependences are described by the power law as well. However, the power law index is different. At low pressure the power law index $\alpha = 2.20\pm 0.06$ for $p=50 ~\rm mm Hg$, $\alpha = 2.04\pm 0.04$ for $p=200~\rm mm Hg$ and  $\alpha = 1.89\pm 0.06$ for $p=360 ~\rm mm Hg$.  This means that if the pressure is kept constant at some $P_{limit}$ the index appears very close to the value obtained in the experiments with Russian GCs.  In general the dependence of the optimal separation power on velocity can be presented as follows
\begin{equation}
   \delta U \left( V \right) = \left\{
    \begin{array}{ll}
      V^{2.6} & \textrm{if } p_{opt} \left( V \right) < p_{limit}
\textrm{,}\\
      V^2     & \textrm{if } p_{opt} \left( V \right) > p_{limit},
   \end{array} \right.
\end{equation}
where $ p_{opt}(V)$ is the optimal pressure.

\subsection{Dependence of the optimal pressure $ p_{opt}$ on the velocity of the rotor}
\begin{figure}[h!]
 \centering
  \includegraphics[width=0.8\textwidth]{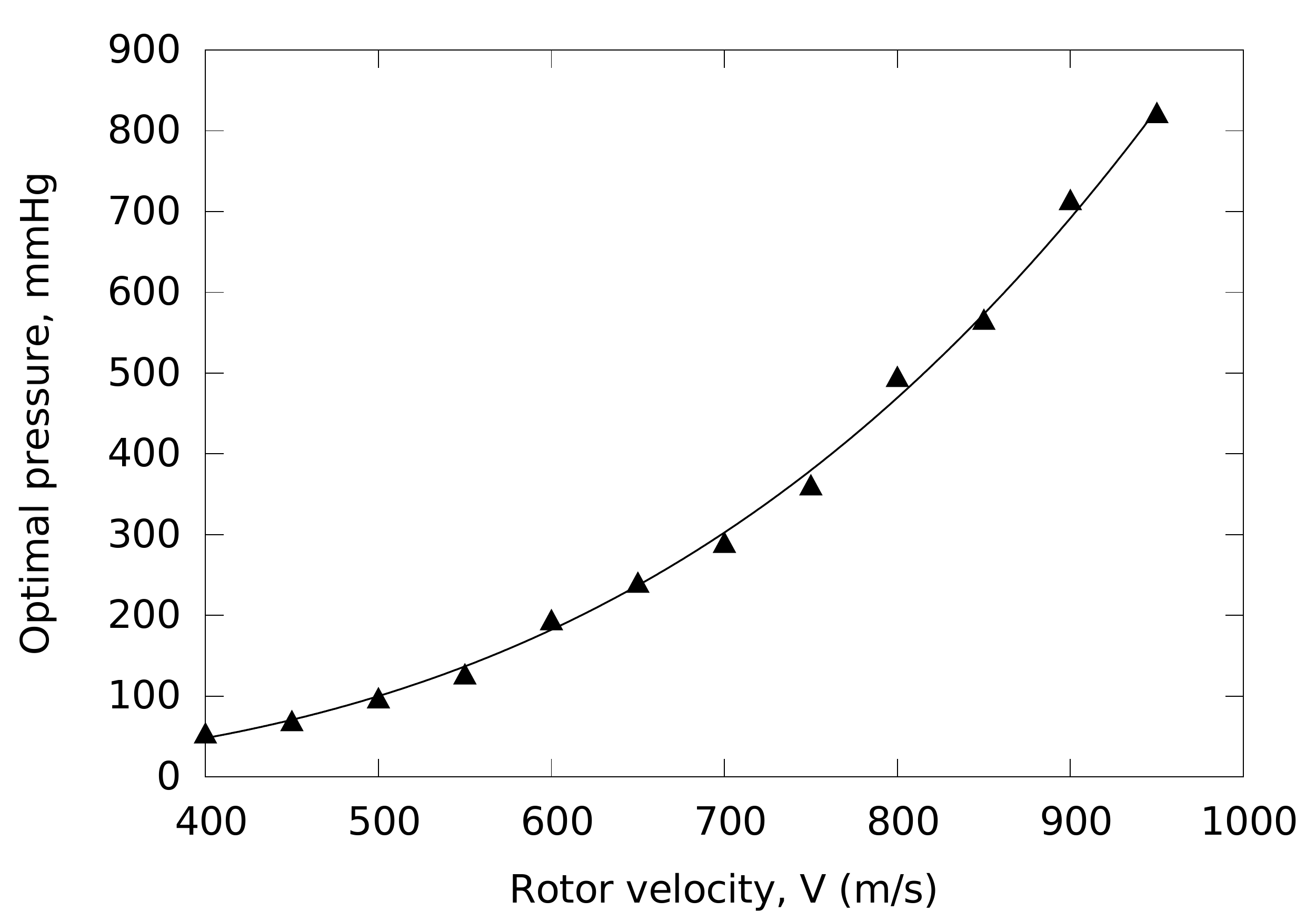}
  \caption{Dependence of the optimal pressure in the gas centrifuge on the velocity of the rotor}
\label{fig3}
\end{figure}

One of the most important parameters of the optimal centrifuge is the pressure at the wall of the rotor. To our knowledge, the pressure in the optimal regime of exploration of the GC has been defined only in the work by \cite{du}. According to this work, the optimal pressure depends on the velocity of the rotor as $V^5$. The numerical simulation gives us unique opportunity to obtain the dependence of the optimal pressure on the rotor velocity. The results of calculations of this dependence  are presented in fig. \ref{fig3}.
This dependence can be approximated by a power law as well. The power law index equals to $3.54\pm 0.08$. This index remarkably differs from $5$  defined in the work  \cite{du}. 

\section{Conclusion}

In the work we have performed numerical simulation of the hydrodynamics and diffusion of isotopes in 2-D model of the Iguassu gas centrifuge. Special procedure of optimization of the gas centrifuge has been performed with the objective to reach maximal separative power at a specified velocity, length and diameter of the rotor.  
The calculations show that the separative power depends on the rotor velocity as $V^{2.6}$. This dependence differs from the dependence $\delta U \sim V^2$ determined experimentally for the Russian centrifuges and differs from the dependence determined for the concurrent centrifuges \cite{bog2016}. This difference is explained by the fact that no limitations on the optimal pressure in the GC was imposed as it is done conventionally in the real centrifuges. If the pressure in the centrifuge is kept constant, the dependence of the  separative power on the rotor velocity becomes close to the experimentally determined.

\section{Acknowledgements}

The work has been performed under the support of  the Ministry of education and science of Russia, grant no. 3.726.2014/K. The computer simulations have been partially performed on the computer cluster of National research nuclear university(MEPhI), Basov farm. This work was also performed within the framework of the Center "Physics of
nonequilibrium atomic systems and composites" supported by MEPhI
Academic Excellence Project (contracts No. 02.a03.21.0005, 27.08.2013).




\begin{thebibliography}{99}

\bibitem{cohen} K.Cohen. { \it The theory of isotope separation as applied to the large-scale production of $U^{235}$}. Ed. by G.M.Murphy, (McGraw-Hill: New York. 1951)
\bibitem{onsager} Wood, H.G.; Morton, J.B. {\it Onsager’s pancake approximation
for the ﬂuid dynamics of a gas centrifuge}. J. Fluid Mech., V. 101. P. 1. (1980)
\bibitem{du}Doneddu, F.; Roblin, P.; Wood, H.G. {\it Optimization studies for
gas centrifuges}. Sep. Sci. Technol., 35 (8): 1207–1221.(2000)
\bibitem{senchenko} Senchenkov, A.P.; Senchenkov, S.A.; Borisevich, V.D.  {\it Gas
centrifuges}. In: ISOTOPES. Properties. Production. Application. Ed.,
Baranov, V. Yu., Fizmatlit: Moscow, Vol. 1: 168–208 (in Russian).(2005)
\bibitem{yatsenko}  Borisevich V.D., Godisov O.N., Yatsenko D.V. {\it Comparison of the circulation efficiency in
gas centrifuges with different geometric
and speed characteristics for uranium
enrichment}. Atomic energy, 116, 5, 363 - 371 (2015)
\bibitem{kemp} Kemp, R.S. {\it Gas centrifuge theory and development: A review
of U.S. programs}. Sci. Glob. Security, 17 (1): 1–19.(2009)
\bibitem{bog2016} Bogovalov S.V., Borman V.D. {\ Dependence of separative power of gas centrifuges on velocity of rotor} Nuclear science and technology. 2016, in press
\bibitem{iguasu}Proc. of the Fifth workshop on Separation Phenomena in Liquids and Gases, Sep. 22-26, 1996. eds. C. Schwab, N. A.S.Rodrigues and H.G.Wood. 
\bibitem{kai}  T. Matsuda, T. Sakurai,H. Takeda, Source-Sink Flow in a Gas Centrifuge, Journal of Fluid Mechanics 69(1), 197–208(1975)
\bibitem{knudsen} E. Von Halle, H.G. Wood, R.A. Lowry, Effect of vacuum core boundary conditions on separation in the gas centrifuge, Nuclear
Technology 62 (6), 325–334(1983)
\bibitem{landau} Landau L. D., Lifshitz E. M., Fluid Mechanics (Butterworth-Heinemann, Oxford, 1987; Fiz-
matlit, Moscow,2001)
\bibitem{godunov} S.K.Godunov, Numerical solution of multidimensional problems of gas dynamics, in russian, 1976, Nauka.
\bibitem{bog13} S.V. Bogovalov, V.D. Borisevich, V.D. Borman, et al. , Verification of numerical codes for modeling of the flow and isotope
separation in gas centrifuges, Computers and Fluids 86, 177-184 (2013)
\bibitem{kislov13}  V.I. Abramov, S.V. Bogovalov, V.D. Borisevich, et al., Verification of software codes for simulation of unsteady flows in a gas
centrifuge, Computational Mathematics and Mathematical Physics 53(6), 789-797(2013)
\bibitem{BOBYQA} M. J. D. Powell, The BOBYQA algorithm for bound constrained optimization without derivatives, Department of Applied Mathematics and Theoretical Physics, Cambridge England, technical report NA2009/06 (2009).
\bibitem{nlopt} Steven G. Johnson, The NLopt nonlinear-optimization package, http://ab-initio.mit.edu/nlopt.


\end{thebibliography}


\end{document}